\documentclass[aps,prb,preprint,superscriptaddress]{revtex4-2}
\usepackage[english]{babel}
\usepackage[utf8x]{inputenc}
\usepackage[T1]{fontenc}
\usepackage{siunitx}
\usepackage{xcolor}
\usepackage{xspace}
\usepackage[version=4]{mhchem}

\usepackage{placeins}
\usepackage{braket}
\usepackage{longtable}
\usepackage{booktabs}
\usepackage{natbib}
\usepackage{bibentry}
\usepackage{amsmath}
\usepackage{amssymb}
\usepackage{graphicx}
\usepackage[colorlinks=true, allcolors=blue,breaklinks=True]{hyperref}

\usepackage{ragged2e}

\begin{document}

\title{Nitrogen-vacancy centre formation via local femto-second laser annealing of diamond}


\author{Davin Yue Ming Peng}
\affiliation{School of Science, RMIT University, Melbourne, VIC 3001, Australia}

\author{Alexander J. Healey}
\email{alexander.healey2@rmit.edu.au}
\affiliation{School of Science, RMIT University, Melbourne, VIC 3001, Australia}

\author{Rebecca Griffin}
\affiliation{School of Science, RMIT University, Melbourne, VIC 3001, Australia}

\author{Benjamin Cumming}
\affiliation{School of Science, RMIT University, Melbourne, VIC 3001, Australia}

\author{Hiroshi Abe}
\affiliation{National Institutes for Quantum Science and Technology: QST, Takasaki, Gunma, 370-1292, Japan}

\author{Takeshi Ohshima}
\affiliation{National Institutes for Quantum Science and Technology: QST, Takasaki, Gunma, 370-1292, Japan}
\affiliation{Department of Materials Science, Tohoku University, Aoba, Sendai, Miyagi, 980-8579, Japan.}

\author{Alastair Stacey}
\affiliation{School of Science, RMIT University, Melbourne, VIC 3001, Australia}

\author{Brant C. Gibson}
\affiliation{School of Science, RMIT University, Melbourne, VIC 3001, Australia}

\author{Brett C. Johnson}
\email{brett.johnson2@rmit.edu.au}
\affiliation{School of Science, RMIT University, Melbourne, VIC 3001, Australia}

\author{Philipp Reineck}
\email{philipp.reineck@rmit.edu.au}
\affiliation{School of Science, RMIT University, Melbourne, VIC 3001, Australia}

\date{\today}

\begin{abstract}
Emerging quantum technologies based on the nitrogen-vacancy (NV) centre in diamond require carefully engineered material with controlled defect density, optimised NV formation processes, and minimal crystal strain. The choice of NV generation technique plays a crucial role in determining the quality and performance of these centres. In this work, we investigate NV centre formation in nitrogen-doped diamond using femtosecond (fs) laser processing. We systematically examine the effect of laser pulse energy on NV production and quality using photoluminescence and optically detected magnetic resonance measurements. We also probe the role of pre-existing lattice defects formed by electron irradiation and consider defect evolution over extended dwell times. Finally, we are able to identify a regime where the main action of the fs-laser is to diffuse rather than create vacancies. This local annealing capability expands the toolkit for tailored NV production and presents opportunities for fine tuning defect populations.
\end{abstract}

\maketitle


\section{Introduction}
\label{sec:intro}

The negatively charged nitrogen-vacancy centre (NV$^{-}$) in diamond is an enabling platform for quantum sensing and quantum information processing\cite{Rovny2024, Pezzagna2021, Schmitt2017,Abobeih2019}. A key requirement for these applications is controlled NV formation. Ion implantation and thermal annealing is often employed for the precise placement of nitrogen atoms in the diamond lattice,\cite{Meijer2006, Schroder2017} however, it is limited by ion straggle and the generation of cascades of displaced atoms around the ion track. The latter can be detrimental to NV spin coherence~\cite{Tetienne2018,Healey2020} and impacts N-to-NV conversion~\cite{Pezzagna2010,Luhmann2019}. Recently, femtosecond (fs) laser writing of colour centres into wide band gap materials has emerged as an alternative.\cite{Wang2022LPR} In this process, localised fs-laser pulses create vacancies in samples already containing substitutional nitrogen, N$\rm _s$.\cite{Chen2017} Subsequent thermal annealing of the entire diamond induces NV formation.\cite{Chen2017,Chen19} Fs-laser writing has been extended to a range of colour centres and host materials,\cite{Wang2022LPR} including group IV defects in diamond,\cite{Cheng2025} and colour centres in silicon carbide\cite{Chen2019,Cheng2025,Castelletto18}, gallium nitride,\cite{Castelletto2021, Guo2025} and hexagonal boron nitride\cite{Hou2018, Gao2021}.

Vacancy creation with a fs-laser is believed to occur through a multi-photon ionisation mechanism, as evidenced by a power-law dependence of NV density on fluence \cite{Kononenko2017}. For example, a 790~nm, 300~fs laser will generate NV centres within a fs-laser energy range of 20-36~nJ per pulse.\cite{Chen2017} Exceeding this energy range results in an irreversible phase transformation into graphitic-like disordered carbon.  Indeed, theoretical simulations have suggested lattice temperatures of 1600~K can be achieved during fs-laser pulses.\cite{griffiths2021} However, the strong power dependence may also present opportunities for controlled defect generation. Furthermore, it was recently shown that multi-pulse fs-laser processing can produce low density ($\approx 60$~ppb) NV ensembles without a subsequent annealing step, implying that the laser itself facilitates vacancy diffusion~\cite{Shimotsuma23}. Coupling these local defect creation and annealing capabilities may allow the resultant defects to be better controlled, with potential benefits for NV spin coherence and yield~\cite{Luhmann2019,Tetienne2018}. 

In this work, we investigate NV centre ensemble formation via fs-laser writing over a range of pulse energies. We begin with an as-received sample containing approximately 1~ppm nitrogen, and identify a regime that efficiently creates NV ensembles without degrading their spin properties. Notably, a bulk annealing step is not required to form the ensembles. We then probe the role of the fs-laser in locally annealing the sample by repeating the experiments on a sample with an existing vacancy density created by electron irradiation. In particular, we establish that pulse energies below the vacancy production threshold can cause vacancy diffusion and NV formation with sufficiently long dwell times. These findings demonstrate the potential for fs-laser writing in fine tuning the creation of NVs in diamond compared to traditional production methods, and may extend to the creation of other defects in diamond and other materials.

\section{Methodology}
\subsection{Diamond samples}

The formation of dense ensembles of NV$^{-}$ centres using fs-laser processing was carried out on optical-grade diamonds from Element Six grown by chemical vapour deposition (CVD). These diamonds have a nominal $\rm N_s$ concentration of 1~ppm and exhibit NV PL in their as-received state without any processing. The manufacturer estimates the NV concentration to be $<$10~ppb.

Two diamonds with distinct processing steps were used to investigate fs-laser defect generation and annealing. The first diamond is an as-received diamond (referred to as the `{\em as-received}' sample), while the other contains a high density of vacancies produced by electron-beam irradiation (referred to as the `{\em e-irrad.}' sample). The electron irradiation was performed before fs-laser writing with a 2~MeV electron beam to a dose of $10^{18}$~cm$^{-2}$. 

\begin{figure}
    \centering
    \includegraphics[width=14 cm]{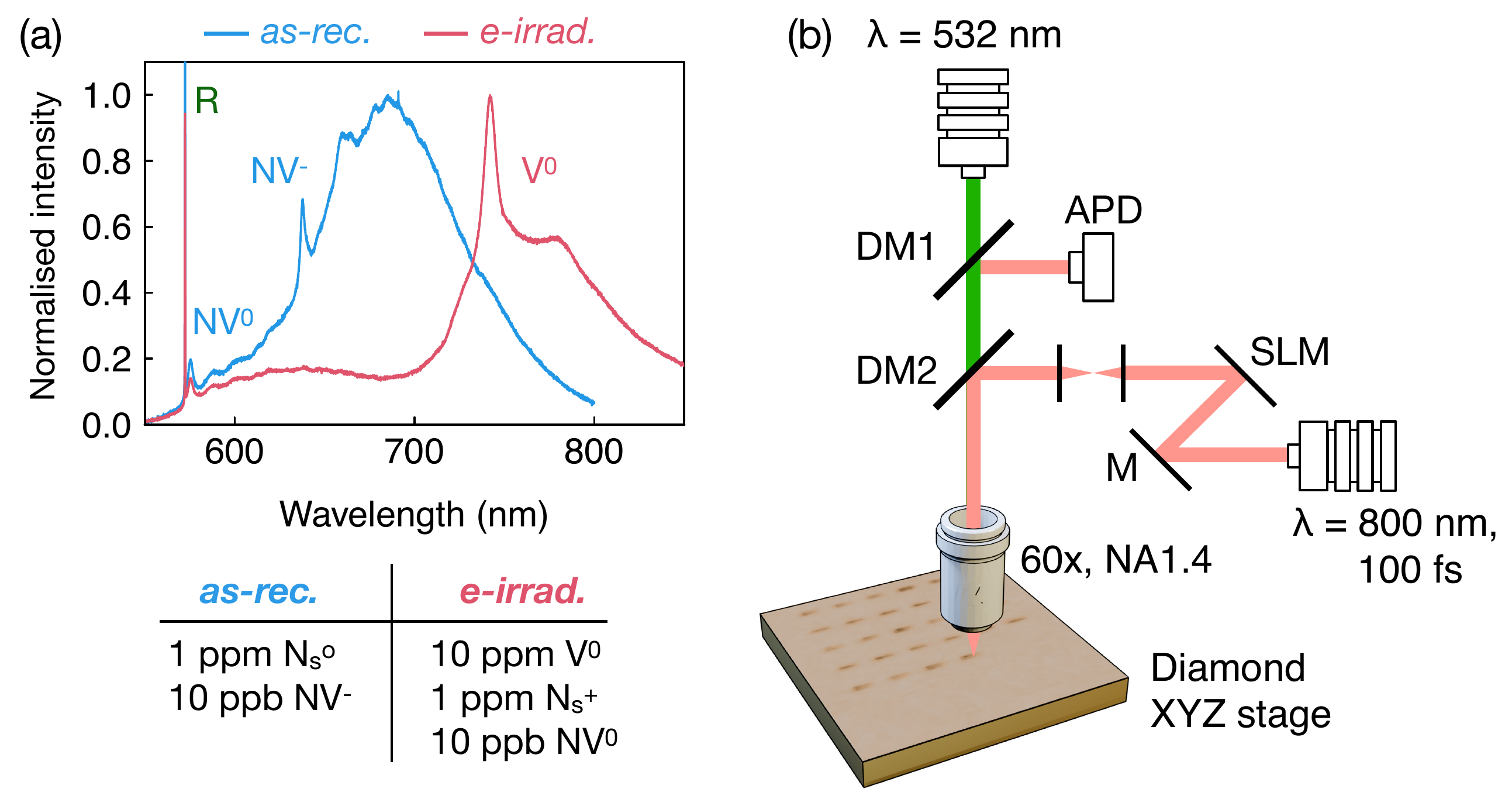}
    \caption{(a) Typical PL spectra from the as-received (blue) and e-irrad. (red) diamonds before fs-laser processing. A table of estimated defect concentrations before fs-laser processing is included below the spectra. (b) A schematic illustration of the combined fs-laser processing and confocal PL imaging setup. The 800~nm fs-laser is directed onto a spatial light modulator (SLM) before steering onto a 60x oil objective with a dichroic mirror (DM2). A 532~nm laser is used to excite the PL of the defects created in this process. PL mapping is performed with scanning mirrors (not shown) independent of the fs-laser. The sample is held on an XYZ piezo-stage.}
    \label{F1}
\end{figure}

Figure~\ref{F1}(a) shows the PL spectra obtained from the as-received and the e-irrad diamonds before fs-laser processing. The NV$^{0}$ and NV$^{-}$ zero phonon lines (ZPL) at 575~nm and 637~nm, respectively, are indicated. The associated phonon sidebands are observed as a broad band centred at about 700~nm. After electron irradiation, the sample shows characteristic PL from neutral vacancy defects (V$^{0}$, also known as the GR1 centre) with a ZPL at 744~nm (as indicated in Fig.~\ref{F1}(a)).  Based on the electron irradiation conditions, we estimate the concentration of V$^{0}$ defects as 10~ppm \cite{Mainwood2000}. V$^{0}$ defects readily accept electrons from nearby substitutional nitrogen atoms ($V^{0} + N_s^{0} \rightarrow V^{-} + N_s^{+}$),\cite{Davies1977} and hence compete with NV centers for electrons so that only the NV$^{0}$ PL signal is observed.

\subsection{Laser writing}
Figure~\ref{F1}(b) shows a schematic setup used to fs-laser process and characterise the diamonds. The fs-laser writing was performed with an Astrella amplified Ti:Sapphire fs-laser, operating at a central wavelength of 800~nm. The repetition rate was set to 5~kHz and the pulse duration was 100~fs. The laser beam was expanded to fill a liquid crystal-based spatial light modulator (HSPDM5120785-PCIe, Boulder Nonlinear Systems). This provided phase-only modulation of the laser beam to compensate for optical aberrations. 

A 7$\times$6 spot array (or elongated lines, 42 spots in total to assess changes statistically) was formed in each sample. Each array was repeated with 16 different pulse energies between 4 and 90~nJ fs-laser energy per pulse. 

\subsection{Optical characterisation}
To characterise the defects generated by fs-laser processing, a custom-built confocal microscope was used, equipped with a spectrometer for PL analysis. A series of optical filters were employed to selectively probe different defect emission characteristics. Optically detected magnetic resonance (ODMR) measurements were also performed by applying microwaves to the diamond sample via a printed circuit board (PCB). An overview of the ODMR measurement principles is provided in the Supporting Information (SI).

\section{Results and discussion}
\subsection{Processing of the {\em as-received} diamond}

\begin{figure}
    \centering
    \includegraphics[width=16 cm]{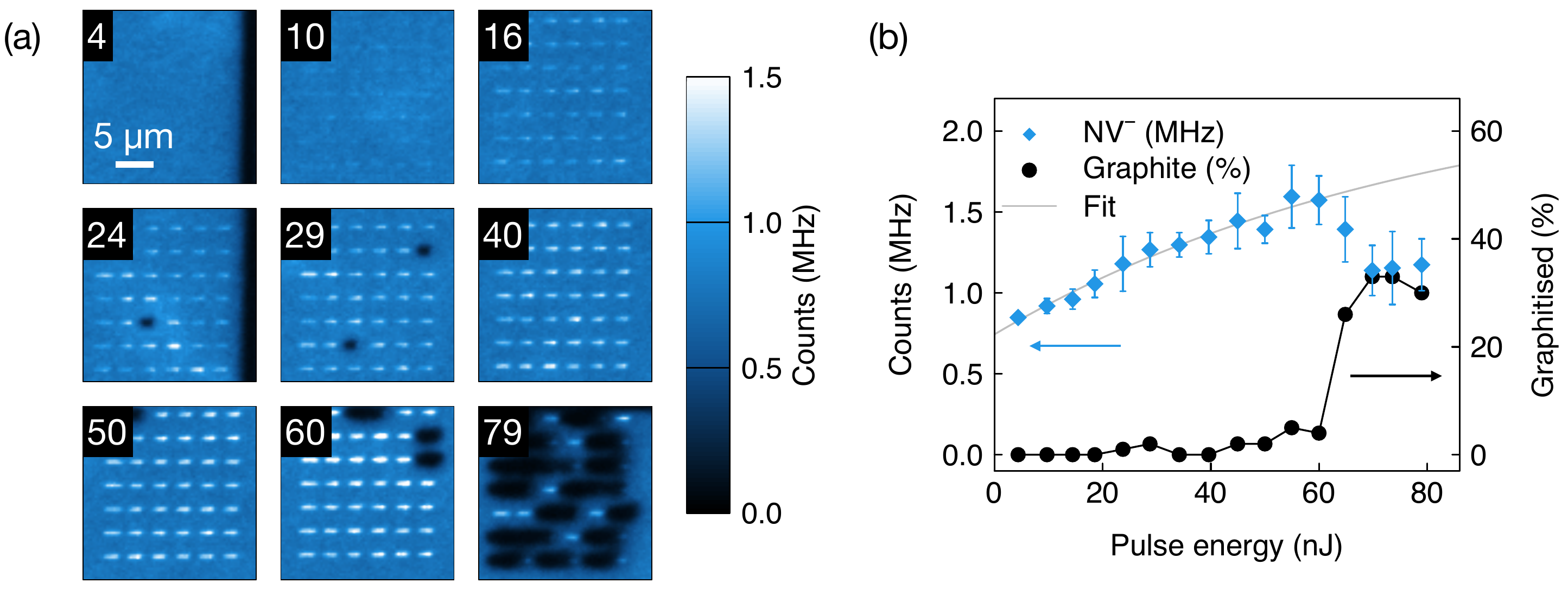}
    \caption{(a) A series of selected confocal PL maps of NV fluorescence (650~nm band pass filter) from the fs-laser processed regions in the as-received diamond. The values in each map indicate the fs-laser pulse energy in units of nJ. (b) The change in NV$^{-}$ PL (blue), and the probability of graphitisation (plotted against the right y-axis).}
    \label{F2}
\end{figure}

We first consider fs-laser processing of the as-received diamond. Figure~\ref{F2}(a) shows selected PL confocal maps of the fs-laser processed {\em as-received} diamond using the indicated laser powers in units of nJ, illustrating the trend across the range of laser powers used. These maps were collected with a 650~nm band pass filter so that both NV$^{-}$ and NV$^{0}$ PL is collected.

As observed in Figure~\ref{F2}(a), a pulse energy of 4~nJ, does not result in any noticeable modification to the PL intensity. At pulse energies between 10 and 60~nJ, the number of visible spots and the brightness of each spot in the laser-written array increases with fs-laser pulse energy, indicating an increasing NV concentration. From around 60~nJ, an increasing number of spots become dark, indicating the increasing probability of transforming the diamond into a graphite-like disordered carbon phase. We refer to this here as graphitisation. 

Figure~\ref{F2}(b) summarises the change in PL intensity as a function of fs-laser pulse energy for the as-received sample. The PL intensity is determined from each spot in the confocal maps that did not graphitise. The error bars are the standard deviation between the PL intensities of individual spots. The solid line is a guide to the eye only. The percentage of spots for each array transforming to disordered graphite is also plotted (against the right axis of Fig.~\ref{F2}(b)). A clear threshold behaviour is observed at a pulse energy of $>60$~nJ which correlates with a drop in the NV PL intensity.

Fs-laser processing in the as-received sample did not result in a clear V$^{0}$ PL signal, suggesting that V$^{0}$ defects were not formed in significant concentrations. The increase in the NV PL was the dominant observation and is notable since no thermal annealing was employed to activate these defects. Interestingly, a weak PL signal could be observed in the graphitised regions. This was only observed with PL confocal measurements using a variable excitation laser wavelength (included in the SI).  

\begin{figure}
    \centering
    \includegraphics[width=15 cm]{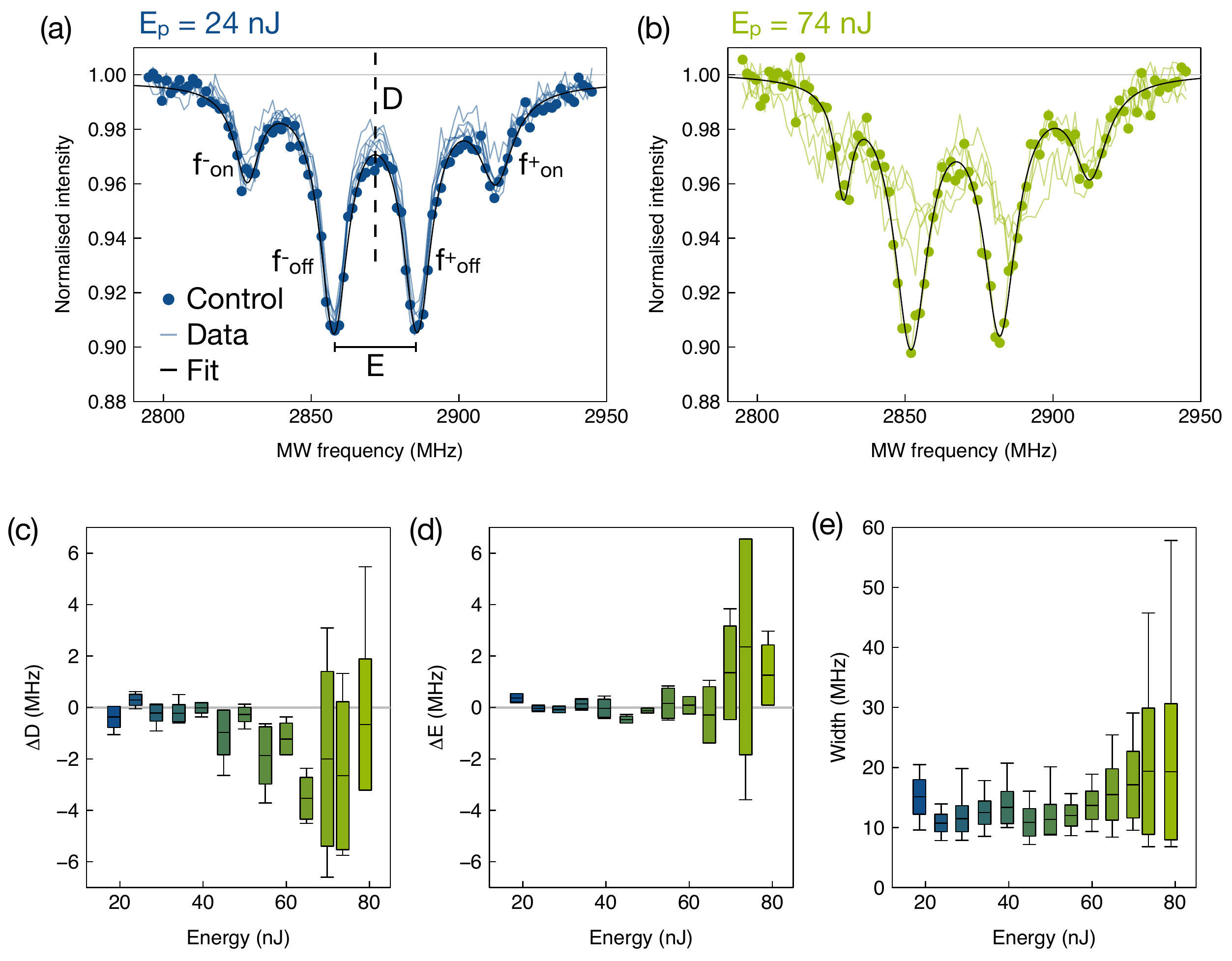}
    \caption{NV$^{-}$ ODMR spectra from fs-laser processed spots with pulse energies of (a) 23.75~nJ and (b) 73.56~nJ. Spectra with solid symbols are background spectra taken in a nearby unprocessed area. The solid black line is a multi-Lorentzian fit. A weak bias magnetic field is applied along one set of NV axes. The change in (c) zero field splitting parameter, $D=(f^+ + f^-)/2$ and (d) the parameter $E= (f^+_{\rm off} - f^-_{\rm off})$ as a function of pulse energy are extracted from these spectra with their definition indicated in (a). (e) The ODMR resonance full-width at half-maximum.}
    \label{F3}
\end{figure}

To assess the quality of the NV$^{-}$ centres at each fs-laser processed spot, we performed ODMR measurements with a confocal microscope. Representative spectra are displayed in Fig.~\ref{F3}(a) and (b) for spots measured in the 24~nJ and 74~nJ array, respectively. During ODMR measurement a small  bias magnetic field was applied along one of the NV axes resulting in four resonances, $f^{\pm}_{\rm on/off}$. These are from the $m_s = 0$ and $m_s = \pm 1$ ground spin states of NVs aligned with the applied bias magnetic field ($f^{\pm}_{\rm on}$) and those at some other angle ($f^{\pm}_{\rm off}$). The centre point of the spectrum, $D$ is the zero field splitting parameter and is sensitive to crystal field splitting and axial strain. The difference between the two central resonances, $f^+_{\rm off}$ and $f^-_{\rm off}$ is related to the local electric field and transverse strain, $E$. 

The spectra represented with solid symbols and a four-Lorentzian fit (solid black trace) in Fig.~\ref{F3}(a) and (b) were recorded in a nearby unprocessed area. The solid colour traces are from a number of fs-laser processed regions (excluding the graphitised spots). The spectra associated with NVs generated with the low energy fs-laser pulse all overlap well, while the spectra acquired on NVs generated at higher laser energy display a high level of splitting, broadening, and variance between spots. This indicates that the fs-laser processing does not significantly affect the NV spin properties if lower laser energies are employed.

The variability of the NV ODMR characteristics as a function of fs-laser pulse energy is summarised in Fig.~\ref{F3}(c) and (d) for $\Delta D$ and $\Delta E$ (the deviations from the control values), respectively, with reference to the background ODMR spectrum. For pulse energies below 40~nJ, $D$, the crystal field splitting parameter, is not significantly affected (Fig.~\ref{F3}(c)). Above this energy, there is a significant increase in the variability of $\Delta D$, with a tendency towards negative shifts up to about 6~MHz. This indicates that strain of up to 0.4~GPa may be present assuming that dD/dP is 14.58~MHz/GPa for hydrostatic strain \cite{Doherty2014PRL}). This value agrees well with Raman scattering measurements of the 1332~cm$^{-1}$ diamond line which shifts by approximately 0.6~cm$^{-1}$ corresponding to a strain of $\sim$0.2~GPa (included in the SI). Raman mapping also shows that the strain extends out from the graphitised zones which likely has an impact on the ODMR spectra. 

The central splitting relative to the background ODMR spectrum,
$\Delta E$ is plotted in Fig.~\ref{F3}(d). $\Delta E$  remains largely unchanged and below ±1 MHz up to fs-laser pulse energies of 60 nJ, above which it becomes significantly more variable and on average shifts to positive values. This transition coincides with the graphitisation threshold identified based on the NV PL results in Fig.\ref{F2} (b). 

At higher pulse energies an increase in the ODMR full width at half maximum (FWHM) is also observed, shown in Fig.~\ref{F3}(e). The FWHM was determined from the average of the multi-Lorentzian fit to the spectrum. A noticeable increase in the variability is again observed for fs-laser pulse energies above 60~nJ. These measurements were performed in a microwave power broadened regime (approximately 10~MHz here), so the values recorded for low pulse energies do not reflect intrinsic line-widths. Low-MW power measurements are provided in the SI and show that the low fs-laser pulse energies do not have an impact on the intrinsic line-width (a $^{13}$C-limited value of around 600 kHz).

These results show that at pulse energies above 60~nJ significant lattice damage is incurred even when a graphitisation event does not take place. For pulse energies below 40~nJ, there is no measurable impact on NV spin properties and the probability of graphitisation is very low.

\subsection{Processing of the {\em E-irrad.} diamond}

\begin{figure}
    \centering
    \includegraphics[width=16 cm]{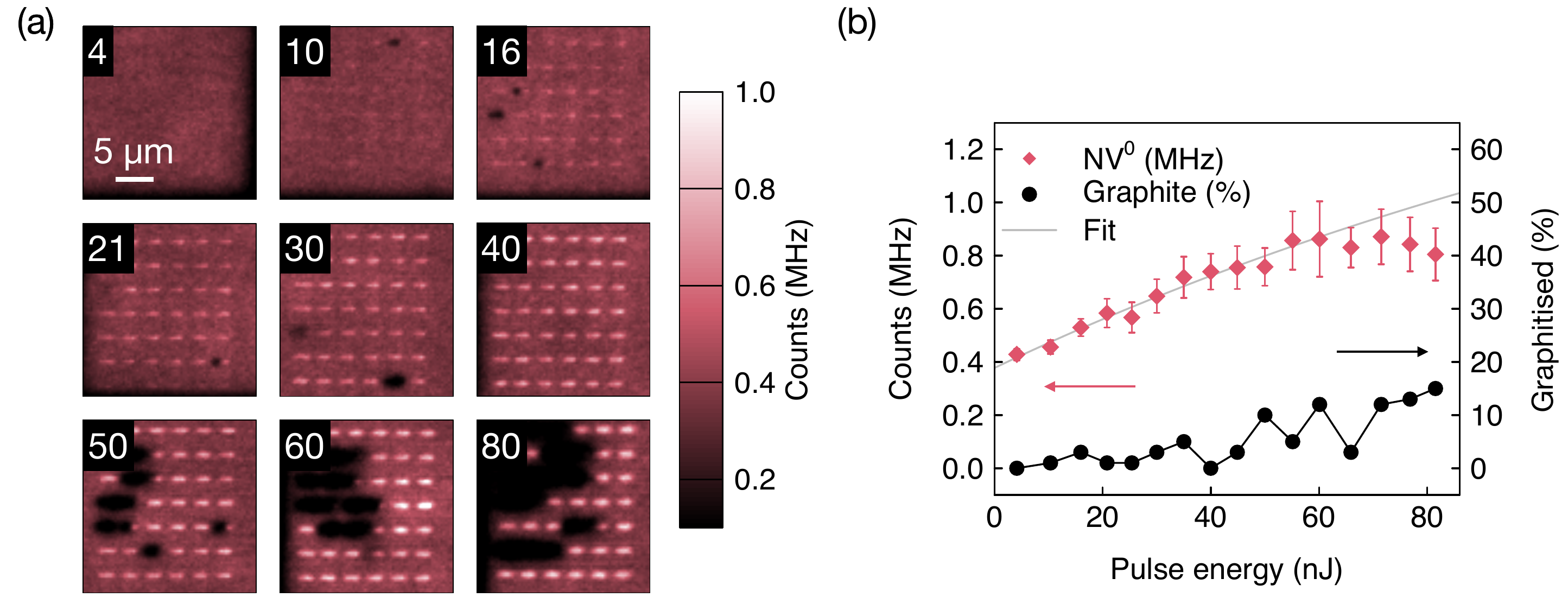}
    \caption{(a) Confocal PL maps of the e-irrad. diamond sample measured with a 650~nm band pass filter. (b) Change in PL intensity (red symbols) and probability of graphitisation as a function of the pulse energy.}
    \label{F4}
\end{figure}

Having established that NV ensembles in as-received diamond can be created by fs-laser writing without producing a measurable increase in single vacancies or lattice strain, we now consider the {\em e-irrad.} diamond which contains a pre-existing vacancy density of approx. 10~ppm  produced by electron irradiation. Figure~\ref{F4}(a) shows selected confocal PL maps after fs-laser processing measured with a band pass filter centred at 650~nm to collect NV PL only and no V$^{0}$ PL. 

Similar to the as-received sample, an increase in the PL intensity is observed with an increase in fs-laser pulse energy. This increase is plotted in Figure~\ref{F4}(b). However, the formation of graphitic material does not display any threshold behaviour as observed with the as-received sample for energies $>60$~nJ (Figure~\ref{F2}(b)). Instead, a gradual increase is observed. We speculate that other defects aid nucleation of graphitisation by acting as local absorption points. Interestingly, unlike in the \textit{as-received} sample, a slight increase in the PL intensity for energies greater than 10~nJ is observed when excited with a 675~nm laser with a 725~nm long pass filter, suggesting a slight increase in the V$^{0}$ concentration (data and discussion in SI). 
 
\begin{figure}
    \centering
    \includegraphics[width=7 cm]{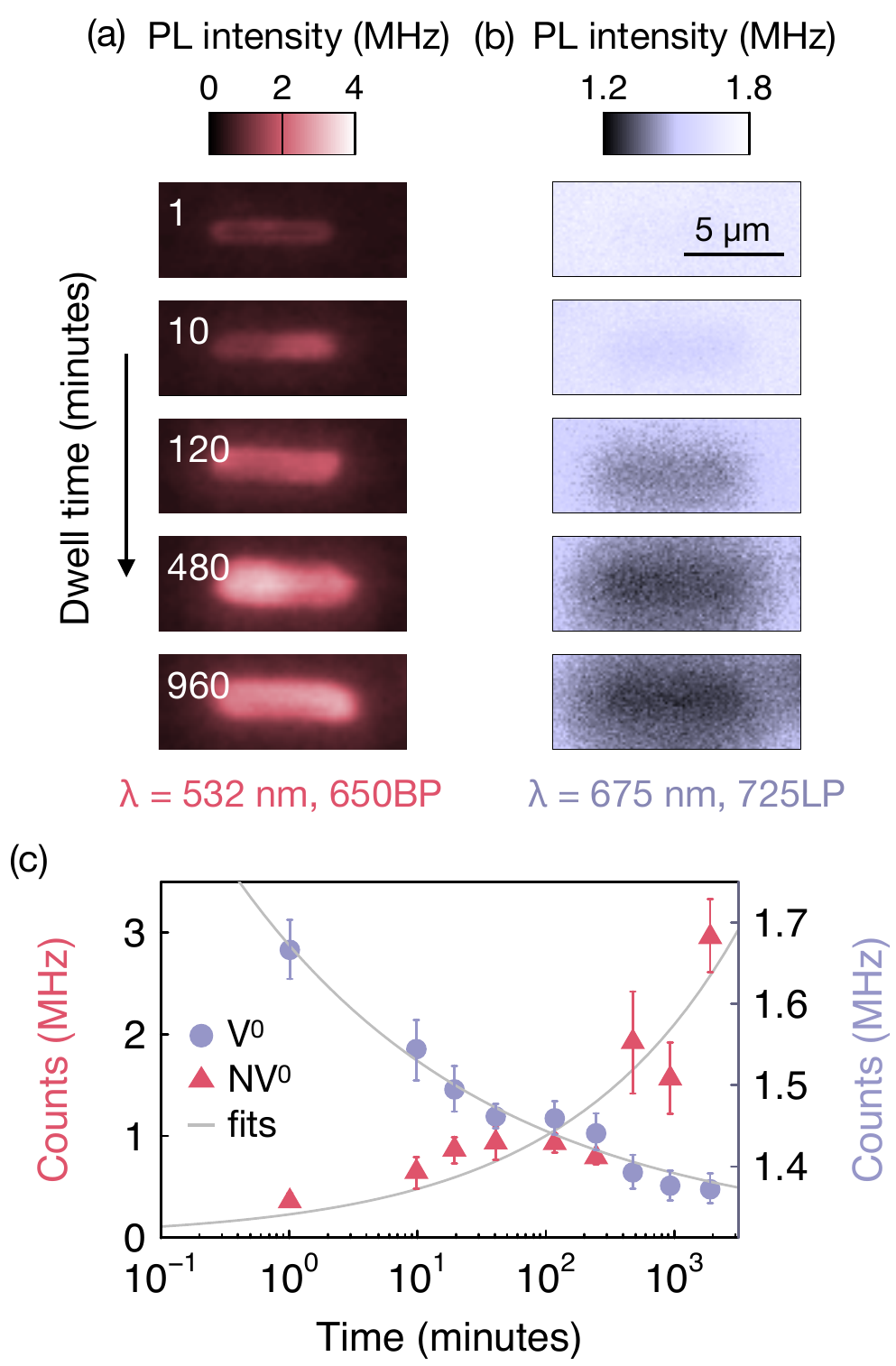}
    \caption{A series of PL maps of 4.5~nJ fs-laser processed regions for the dwell times indicated (in minutes) in the e-irrad. sample. The maps were collected with (a) 532~nm excitation and a 650 band pass filter to collect NV PL and (b) a 675~nm excitation and 725 long pass filter to collect V$^{0}$ PL. (c) The PL intensity versus dwell time. The data points represent the PL intensity averaged over the processed spot and the error bars are the standard deviation of these values.}
    \label{F5}
\end{figure}

To investigate the limits on NV conversion and assess whether the fs-laser locally anneals the diamond, we considered the dwell time in the \textit{e-irrad.} diamond. A low pulse energy of 4.5~nJ was chosen to minimise the probability of graphitisation and excess vacancy production. Fig.~\ref{F5} presents confocal maps acquired with (a) 532~nm excitation and a 650~nm band pass filter (to collect NV PL) and (b) 675~nm excitation with a 725~nm long pass filter (for V$^0$) for various fs-laser dwell times. Although, NV$^0$ production was not clearly apparent in Fig.~\ref{F4} for this laser power, we see a clear increase in NV PL up to the longest dwell time considered, 32 hours. As observed in Fig.~\ref{F5}(b) this is accompanied by a drop in the V$^{0}$ PL which becomes evident after 10~minutes. A control experiment repeated in the \textit{as-received} diamond confirmed that no NVs are produced, indicating that we do not create vacancies at this pulse energy (see SI).

Figure~\ref{F5}(c) summarises these trends and shows the monotonic increase in NV$^0$ and corresponding decrease in V$^{0}$ PL intensity over time. The relative magnitudes of NV$^0$ PL increase and V$^{0}$ depletion should be interpreted with caution due to the differing brightness of the respective defects, their initial abundance, and changes in excitation optics for the two sets of measurements. For example, the PL intensity for a single  V$^{0}$ defect is expected to be orders of magnitude less than that for NV.\cite{Jelezko2006} Therefore, only a small change in the V$^{0}$ PL is expected. In any case, the observed trends are consistent with a local annealing effect, where vacancies diffuse and combine with existing substitutional nitrogen to form NV centres. 

We also note that the short-time decrease in V$^0$ PL is proportionally larger than the NV$^0$ increase, which could indicate multiple vacancy annihilation processes that proceed on different timescales. For instance, since the estimated initial vacancy density exceeds N$_s$ by an order of magnitude, divacancy formation between nearby neutral vacancies is probably significant and perhaps dominant at short times. This may also explain the apparent greater extent of the area affected by the fs-laser appears larger in the V$^{0}$ PL maps (see also Fig.~S5). 

To properly compare the efficacy of fs-laser writing for scalable NV production with traditional irradiate-anneal ensemble fabrication, the N-to-NV conversion efficiency must be considered. Unfortunately, it is difficult to accurately quantify the local N-to-NV conversion in our samples because the focal depth of both the NV-creation laser and confocal collection optics are not precisely known. We can, however, note that the NV PL is observed to increase by a maximum of approximately a factor of two in the both {\em as-received}  and \textit{e-irradiated} diamonds in the single-pulse arrays. Given the relative proportion of NV to $N_S$ in as-grown CVD diamond is typically 0.1-1\%~\cite{Edmonds2012}, we can estimate that the PL increase may be associated with a NV density increase of this order. This is much lower than that achieved using electron irradiation and high temperature annealing (of order 10\% conversion efficiency for bulk diamond). This being said, our dwell time measurements (Fig.~\ref{F5}, where we see an order of magnitude increase in NV$^0$ PL) do suggest that greater conversions can be achieved through prolonged fs-laser processing. The significant length of these dwell times (10s of hours) may be shortened if the fs-laser power is further tuned. Further work is planned to test if this can translate to creating a NV$^{-}$ ensemble with a higher N-to-NV conversion than is achievable through electron irradiation and annealing while maintaining NV spin properties. If this step can be demonstrated, then fs laser-induced NV creation may be viable for activating NVs in individual nano or microdiamond particles. 

\section{Conclusion}
We have shown that fs-laser writing alone can successfully create NV ensembles without compromising NV spin properties. While not as scalable as electron irradiation and annealing of bulk material, the precise defect localisation afforded by laser writing presents advantages for incorporating NV ensembles into existing photonic structures. Importantly, we have shown that as well as creating vacancies at high pulse energies, fs-laser writing locally anneals the diamond even at low energies. This property may hold particular relevance for NV production near surfaces, where oxidation and graphitisation of diamond under high temperature annealing conditions can lead to significant loss of material. 

We have also shown that the high non-linearity in the defect production process with fs-laser parameters allows for fine-tuned defect creation and evolution. As an example, the results presented in Fig.~\ref{F5} show that we can identify a regime where vacancies diffuse but are not created. This, coupled with the technique's operation at ambient lab conditions, presents intriguing opportunities for optimising colour centre creation. For instance, an appealing direction for future work is to combine fs-laser defect creation with control over defect charge states during the local annealing~\cite{Luhmann2019}. By slowly releasing and consuming vacancies it may be easier to keep the charged proportion near 100\% throughout and avoid vacancy clustering, with benefits for both ensemble and single defect creation and operation. Indeed for high pulse energies we observed an increased V$^0$ PL signal in the \textit{e-irrad.} sample but not the \textit{as-received}, suggesting that the majority of vacancies produced are negatively charged even with only 1~ppm N.

\section*{Acknowledgments}
This work was supported by the Australian Research Council (ARC) through grant DP220102518. P.R. acknowledges support through an RMIT University Vice-Chancellor’s Research Fellowship and ARC DECRA fellowship (DE200100279). This work was also partially supported by the Australian Government Department of Defence through the Next Generation Technologies Fund (NGTF). This program is now closed. Extant NGTF activities are being managed under the Advanced Strategic Capabilities Accelerator.

\bibliography{library}

\end{document}


\title{Supporting information: Nitrogen-vacancy centre formation via local femto-second laser annealing of diamond}


\author{Davin Yue Ming Peng}
\affiliation{School of Science, RMIT University, Melbourne, VIC 3001, Australia}

\author{Alexander J. Healey}
\email{alexander.healey2@rmit.edu.au}
\affiliation{School of Science, RMIT University, Melbourne, VIC 3001, Australia}

\author{Rebecca Griffin}
\affiliation{School of Science, RMIT University, Melbourne, VIC 3001, Australia}

\author{Benjamin Cumming}
\affiliation{School of Science, RMIT University, Melbourne, VIC 3001, Australia}

\author{Hiroshi Abe}
\affiliation{National Institutes for Quantum and Radiological Science and Technology, Takasaki, Gunma, 370-1292, Japan}

\author{Takeshi Ohshima}
\affiliation{National Institutes for Quantum and Radiological Science and Technology, Takasaki, Gunma, 370-1292, Japan}
\affiliation{Department of Materials Science, Tohoku University, Aoba, Sendai, Miyagi, 980-8579, Japan.}


\author{Alastair Stacey}
\affiliation{School of Science, RMIT University, Melbourne, VIC 3001, Australia}

\author{Brant C. Gibson}
\affiliation{School of Science, RMIT University, Melbourne, VIC 3001, Australia}

\author{Brett C. Johnson}
\email{brett.johnson2@rmit.edu.au}
\affiliation{School of Science, RMIT University, Melbourne, VIC 3001, Australia}

\author{Philipp Reineck}
\email{philipp.reineck@rmit.edu.au}
\affiliation{School of Science, RMIT University, Melbourne, VIC 3001, Australia}




\maketitle

\section{Fs-laser energies}
\begin{figure}
    \centering
    \includegraphics[width=12 cm]{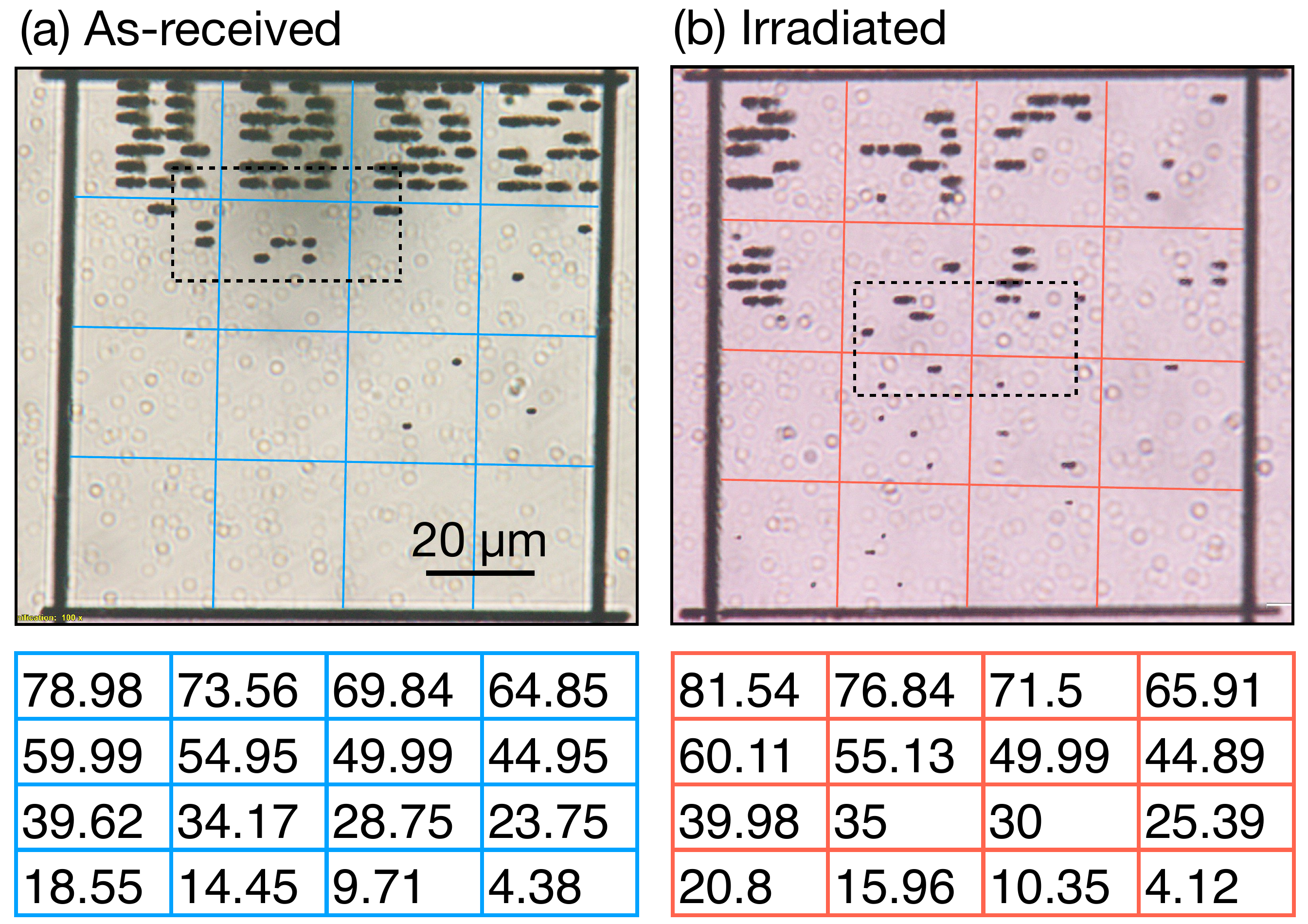}
    \caption{Bright field images of (a) {\em as-received} and (b) the {\em e-irrad.}~diamonds studied in this work. The blue and red grid lines denote the regions in which an array of laser irradiated regions were written. Below each image are the corresponding fs-laser energies in (nJ/pulse). The dashed squares show regions in which Raman maps were recorded (shown below in Fig.~\ref{S4}).}
    \label{S1}
\end{figure}

Bright field images of the fs-laser processed regions in the {\em as-received} and {\em e-irrad.}~diamonds are presented in Fig.~\ref{S1}. A table of the fs-laser energies are given under each image in units of nJ/pulse. The graphitised regions can be clearly observed as dark lines. As noted in the main text, the probability of graphitization has a threshold type behaviour for the {\em as-received} diamond but, when pre-existing damage is present, graphitisation can occur even at low fs-laser energies, as observed in the {\em e-irrad.}~diamond. Selected PL maps of these arrays are shown in Fig.~2 and 3 of the main text.

\section{ODMR linewidths}
\begin{figure}
    \centering
    \includegraphics[width=10 cm]{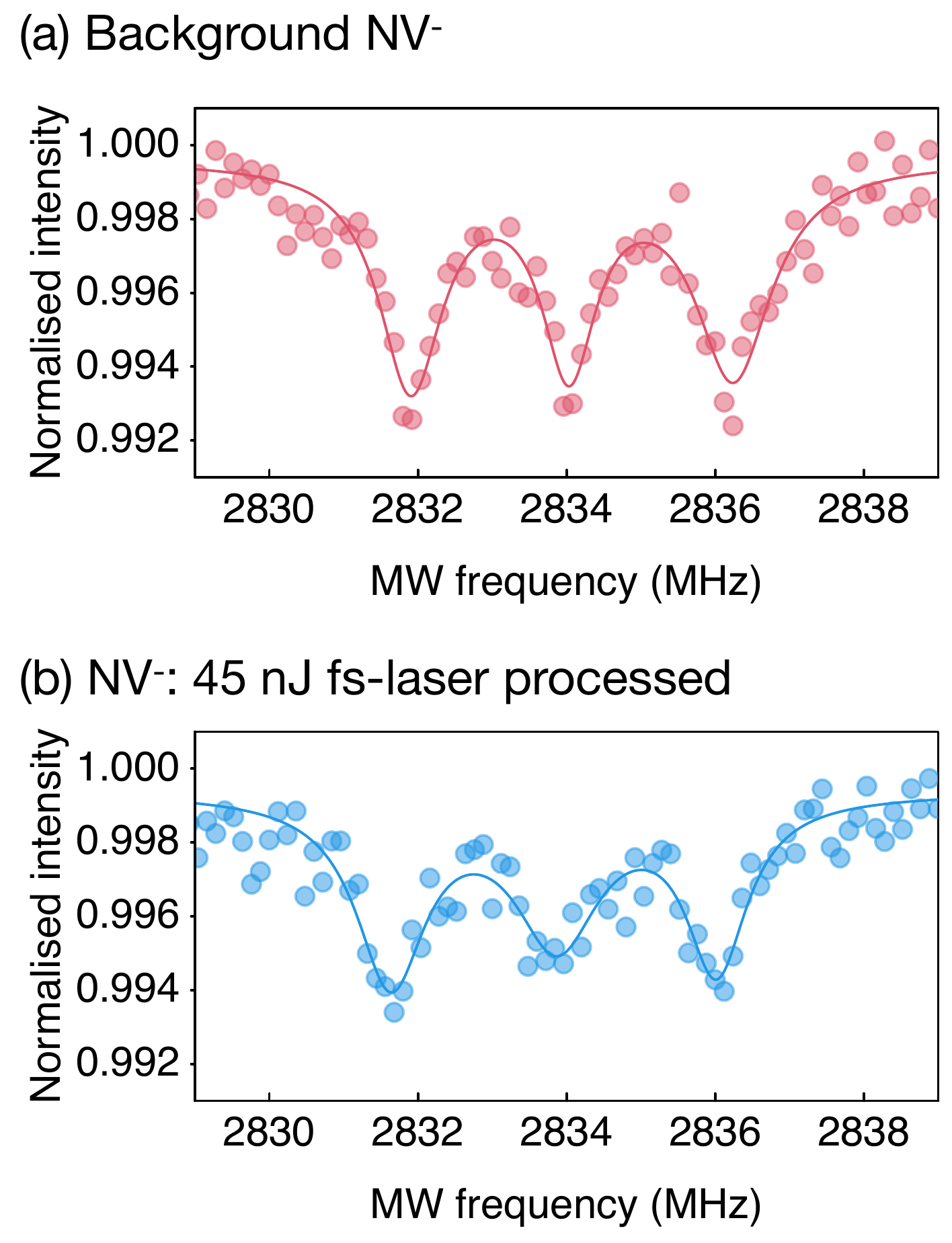}
    \caption{ODMR from NV in (a) the unprocessed region and (b) a 45 nJ fs-laser processed region. A small shift in central resonance frequencies likely reflects a gradient in background magnetic field. Solid lines are a three Lorentzian fit to the data.}
    \label{S2}
\end{figure}

For NV centres formed with a fs-laser we establish that the ODMR linewidth has a fs-laser dependence which we associate with the strain generated by defects in the crystal. This is most prominent when graphitisation occurs which has a different density to diamond and thus generates more strain. 

The ODMR spectra in Fig.~3 of the main text were recorded with a high MW power so that values much larger than the intrinsic linewidth were recorded. In Fig.~\ref{S2} we present ODMR spectra obtained with weak MW driving to avoid power broadening. In Fig.~\ref{S2} we show a $T_2^*$-limited linewidth from a region without fs-laser processing (Fig.~\ref{S2}(a)) compared to a 45~nJ spot (Fig.~\ref{S2}(b)). The pulse energy does not appear to broaden the linewidth significantly, maintaining the $^{13}$C-limited linewidth of around 600~kHz.

\section{Raman spectroscopy}

\begin{figure}
    \centering
    \includegraphics[width=12 cm]{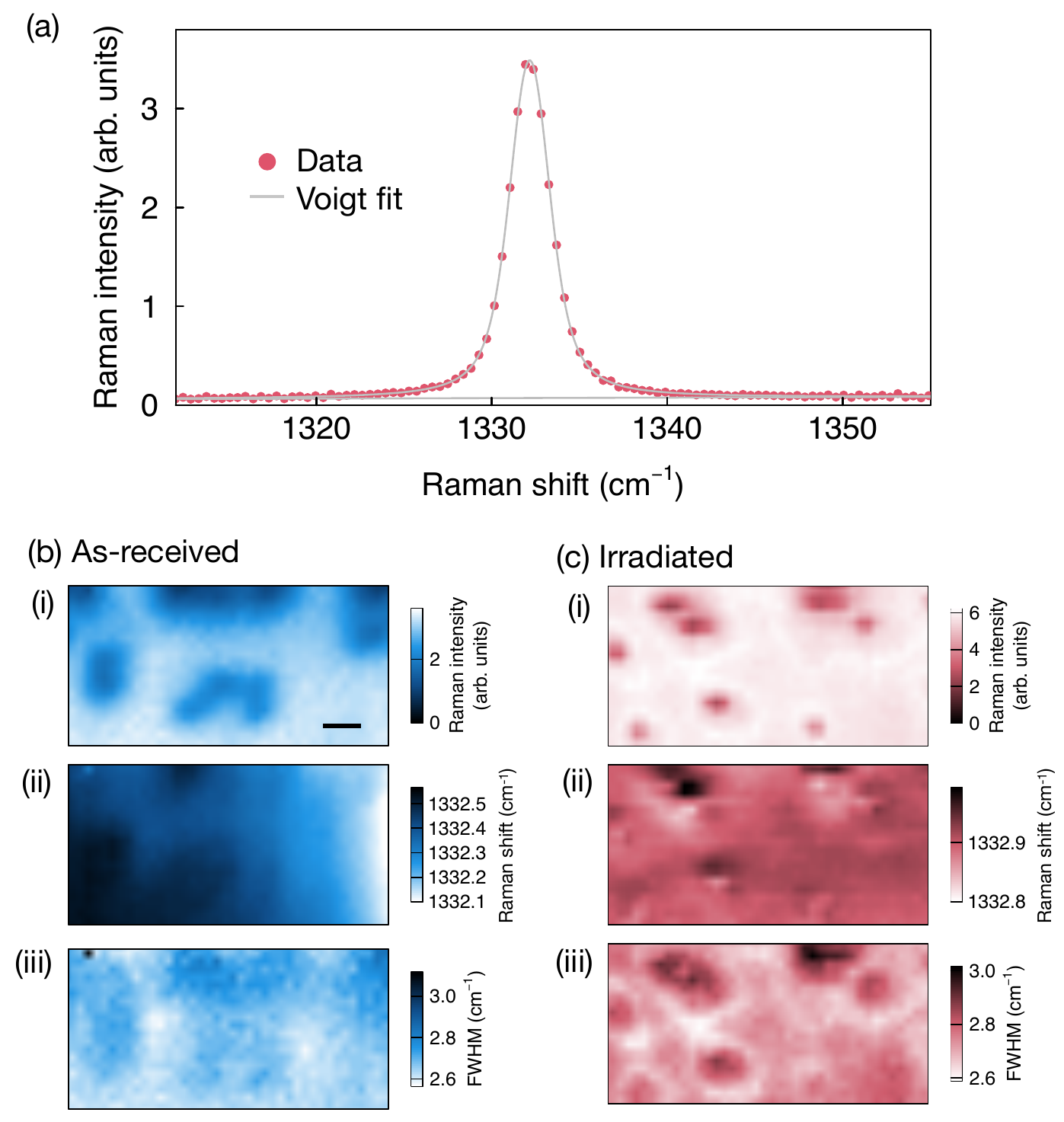}
    \caption{(a) A standard Raman spectrum of a diamond excited with a 532~nm laser, showing the first-order peak at 1332~cm$^{-1}$ fit with a Voigt function. Raman maps of the 1332~cm$^{-1}$ peak intensity (i), Raman shift (ii) and FWHM (iii) are shown for (b) the as-received and (c) the irradiated diamonds.}
    \label{S3}
\end{figure}

Figure~\ref{S3} shows a standard Raman spectrum of diamond. A first-order peak at 1332~cm$^{-1}$ can be observed. This Raman line is a sensitive gauge of disorder in the diamond lattice. Crystal quality, strain, phonon scattering from defects and phase transformations can cause measurable changes in this Raman line. Here, we map changes in Raman intensity, Raman shift and FWHM in the regions indicated with the dashed box in Fig.~\ref{S1}(a) and (b). These maps are displayed in Fig.~\ref{S3} (b) and (c) for the as-received and irradiated diamonds, respectively.

Where the diamond has transformed to a disorder graphitic-like phase a large reduction in the Raman intensity is observed due to the decrease in the diamond volume being probed and possibly phonon scattering. The Raman shift is observed to shift to higher wavenumber especially in the irradiated sample but by a relatively small amount of 0.4 and 0.2~cm$^{-1}$ in the as-received and irradiated samples respectively. If we attribute this shift solely to strain  
frequency shift of 0.15 cm corresponds to the strain of 0.01\%-0.03\%. 

\section{Laser dwell time measurements on as-received sample}
To confirm that a pulse energy of 4.5~nJ does not produce vacancies we repeated the fs-laser dwell time measurement on the as-received sample. Dwell times of 30, 60, 120, 240, 480, 960, and 1920 minutes were considered to match the high end of the e-irrad. series.

\begin{figure}
    \centering
    \includegraphics[scale=1]{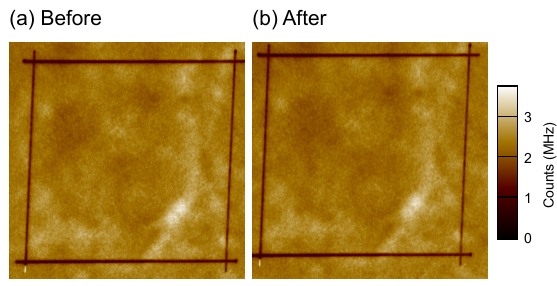}
    \caption{PL maps recorded before (a) and after (b) a fs-laser dwell time series in the as-received diamond. The fs-laser parameters are equivalent to those for the results presented in Fig.~5 of the main text.}
    \label{fig:S_dwelltimeasreceived}
\end{figure}

Confocal maps collected using 500~$\mu$W 532~nm CW excitation (570~nm LP filter) before and after the fs-laser irradiation are shown in Fig.~\ref{fig:S_dwelltimeasreceived}. No local increase in PL is observed above the existing NV background, indicating we do not produce excess NV$^0$, NV$^-$, or V$^0$ in this experiment. Comparing with the results presented in Fig.~5 of the main text, we conclude that fs-laser irradiation at 4.5~nJ does not produce lattice vacancies but does locally anneal the diamond. 

\section{Defects during fs-laser dwell}

\begin{figure}
    \centering
    \includegraphics[width=17 cm]{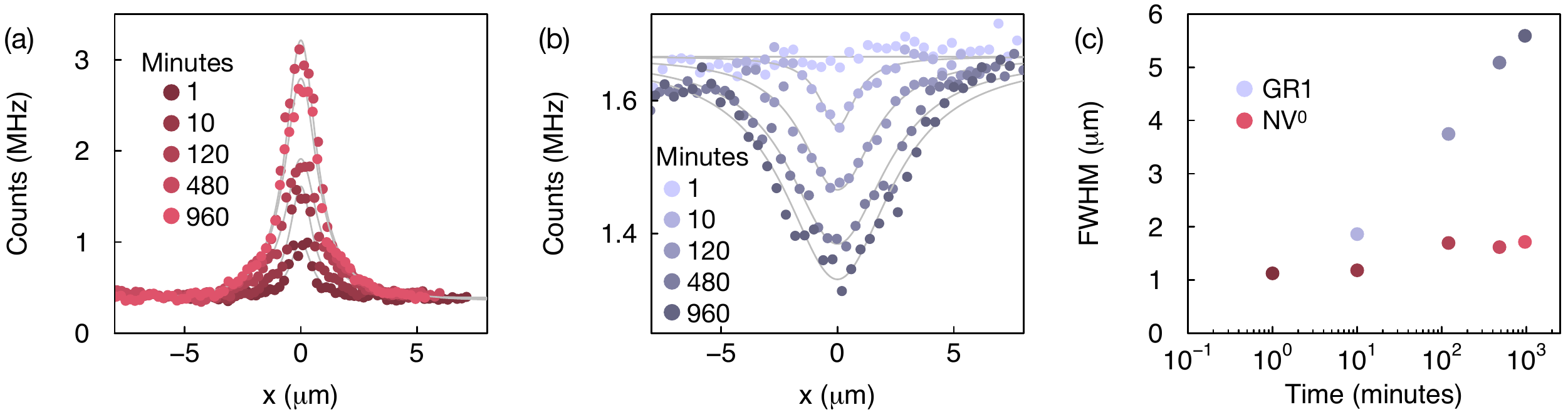}
    \caption{Vertical line-cuts extracted from the confocal maps shown in Fig.~5 of the main text.}
    \label{S5}
\end{figure}

NV$^0$ and V$^0$ PL maps were presented in Fig.~5 of the main text showing the change in PL as a function of the fs-laser dwell time. In Fig.~\ref{S5} we present vertical line cuts of these confocal maps to further highlight the change in defect distributions as a function of fs-laser dwell time.  For NV$^0$ in Fig.~\ref{S5}(a), the fs-laser processed area does not change appreciably over time. Interestingly, the area of of V$^0$ PL depletion increases noticeably with dwell time. This may suggest diffusion of vacancies into the fs-laser processed region once these defects are depleted via laser annealing. 

\section{Photoluminescence from fs-laser processed diamond}

\begin{figure}
    \centering
    \includegraphics[width=12 cm]{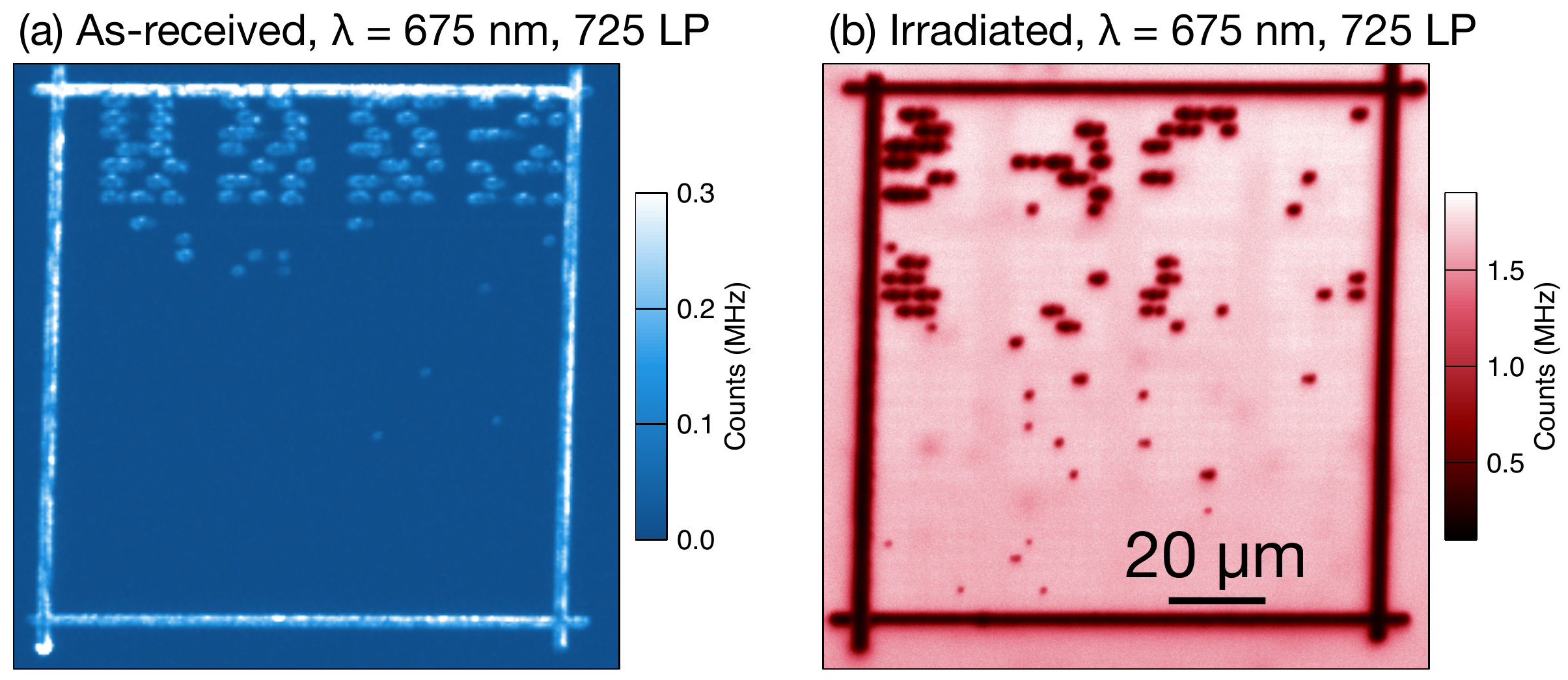}
    \caption{Confocal images of (a) the as-received and (b) the irradiated diamonds. Confocal maps were collected with a 0.5~mW 675~nm pulsed laser and the PL signal was collected via a 725~nm long pass filter.
}
    \label{S4}
\end{figure}

Confocal images of the diamonds studied in this work were presented in Fig.~2 and 4 of the main text. In Fig.~\ref{S4} we present similar maps but using a 0.5~mW 675~nm laser excitation. The PL signal was collected via a 725~nm long pass filter. With this excitation wavelength the NV centre is not excited. Only the signal from the V$^{0}$ signal with a zero phonon line at 744~nm is excited. These maps thus represent the V$^{0}$ distribution in our diamonds. The $4\times 4 $ grid of fs-laser processed arrays is bounded by a laser ablated square. The as-received diamond confocal map in Fig.~\ref{S4}(a) shows essentially no V$^{0}$ PL signal in the fs-laser regions except where graphitisation has occurred. The graphitic regions exhibit a relatively weak signal and will be discussed further below. In the e-irradiated sample the fs-laser processed regions can be observed with a slightly greater V$^{0}$ intensity than the background. As observed in Fig.~4 of the main text, these regions also have an increase in the NV$^{0}$ concentration. The signal observed in Fig.~\ref{S4}(b) is not from NV$^{0}$ but from V$^{0}$.

\begin{figure}
    \centering
    \includegraphics[width=15 cm]{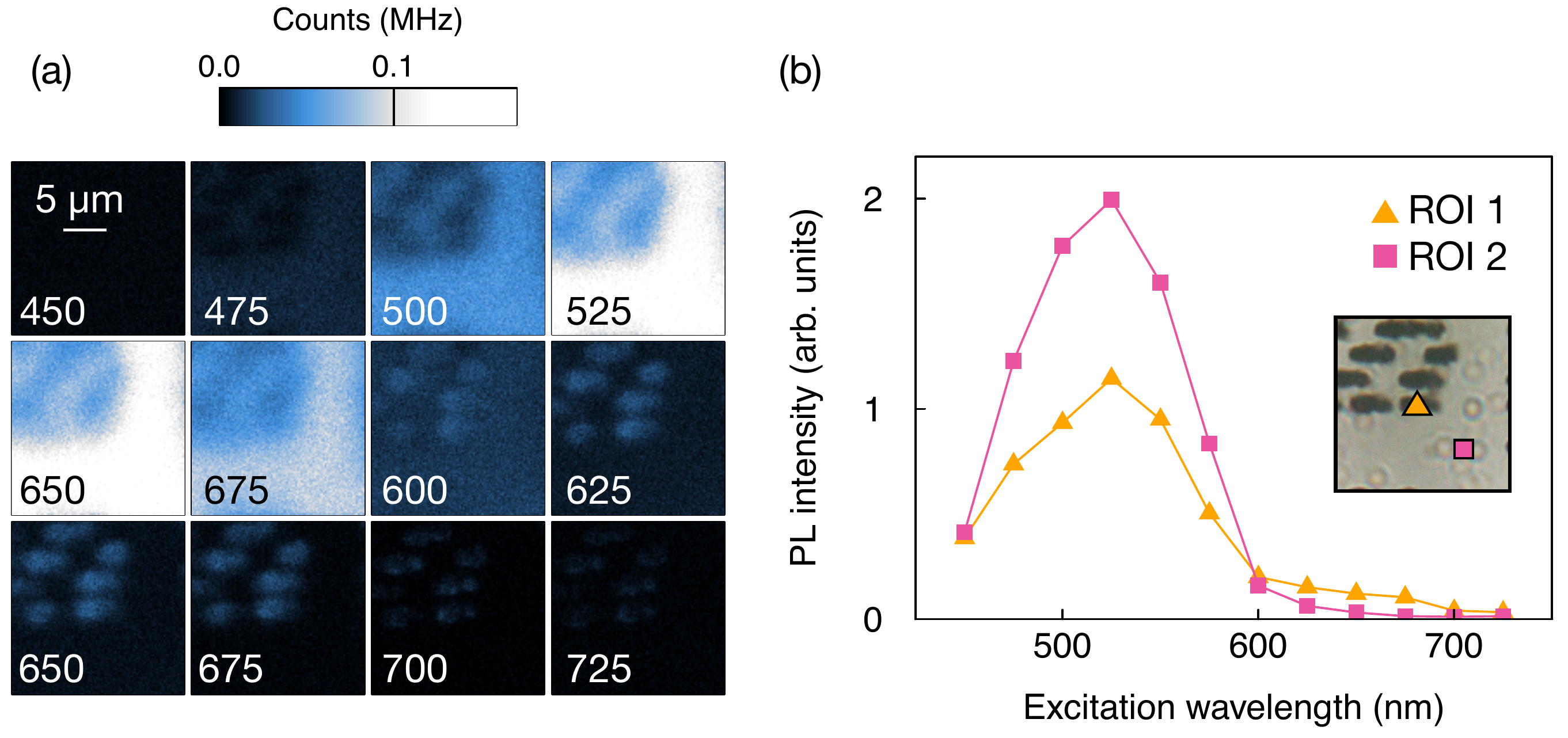}
    \caption{(a) Confocal maps ($20 \times 20 \;\rm \mu m^2$) of the as-received diamond collected with the excitation wavelength indicated (nm) and a 750 nm long pass filter. (b) PL intensity as a function of excitation wavelength in the graphitised (orange triangle) and background (pink squares) regions of interest marked in the bright field image in the inset. 
}
    \label{S6}
\end{figure}

In the as-received diamond we saw that the PL was dominated by NV$^{-}$, with no clear evidence for the presence of V$^{0}$. The lack of a V$^{0}$ signal is clear in Fig.~\ref{S4}. In Fig.~\ref{S6}, we vary the excitation wavelength to preferentially excite various defects in the diamond. Fig.~\ref{S6}(a) shows a series of confocal maps from the as-received diamond collected with a range of excitation wavelengths in the range 450-725~nm. Fig.~\ref{S6}(b) summarises the PL intensity in the graphitised (orange triangle) and background (pink squares) regions in the 73.56~nJ array. The PL intensity peaks around 525~nm, following the expected NV$^{-}$ absorption lineshape. For excitation wavelengths above 600~nm, the graphitised region becomes brighter than the background NV PL. Although a clear V$^{0}$ spectrum could not be obtained, we speculate that this increase in PL is from  a small concentration of V$^{0}$ (the graphite itself is not expected to emit PL). 

One way to reconcile these results is to consider the V$^-$ population, which is not measured in our PL scans~\cite{Davies1977}. The small amount of extra vacancies that are produced tend to be negatively charged in the as-received diamond due to the abundance of uncompensated nitrogen donors, while in the \textit{e-irrad.} sample excess vacancies will be in the neutral charge state since there are already enough vacancies to compensate the nitrogen. Graphitic carbon in the as-received sample introduces acceptor states that may act similarly to depopulate V$^-$ in favour of V$^0$.



\bibliography{library}